\documentclass[12pt,tightenlines,eqsecnum,floats,aps,amsmath,amssymb,nofootinbib,prd,showpacs]{revtex4}

\usepackage{setspace}
\usepackage{amsmath,amssymb,amsfonts,amsthm,amscd}
\usepackage{setspace}
\usepackage{graphicx}
\usepackage{enumerate} % advanced enumerate environment
\usepackage{colordvi} % for color text

\def\be{\begin{equation}} % Makes equation + all the formatting and labeling
\def\ee{\end{equation}}
\def\ba{\begin{eqnarray}}
\def\ea{\end{eqnarray}}
\def\bi{\begin{itemize}}
\def\ei{\end{itemize}}
\def\bnum{\begin{enumerate}}
\def\enum{\end{enumerate}}

\def\ep{\epsilon}
\def\Ga{\Gamma}
\def\E{\tilde{E}}
\def\N{\utw{N}}
\def\Ct{\tilde{C}}
\def\Kbart{\tilde{P}}
\def\Pt{\tilde{P}}
\def\Dt{\tilde{D}}

\def\C{{C}}
\def\Kbar{{P}}
\def\P{{P}}
\def\D{{D}}
\def\CE{\mathcal{E}}
\def\q{\mathring{q}}

\def\f{\frac}

\def\utw#1{\rlap{\lower1ex\hbox{$\sim$}}#1{}} %Improve?
\usepackage{colordvi}
\newcounter{mnotecount}[section]

\newcommand{\comment}[1]{}
\begin{document}

%\preprint{\vbox{\baselineskip=12pt \rightline{IGC-11/02-05}}}
%\rightline{gr-qc/yymmnnn}
%\rightline{NSF-KITP-07-05 }}}

\title{A Hamiltonian Formulation of the BKL Conjecture}

\author{Abhay Ashtekar${}^1$}\email{ashtekar@gravity.psu.edu}
\author{Adam Henderson${}^1$}\email{adh195@psu.edu}
\author{David Sloan${}^{1,2}$}\email{sloan@gravity.psu.edu}
\affiliation{${}^1$ Institute for Gravitation and the Cosmos,
Physics Department, Penn
State, University Park, PA 16802, U.S.A.\\
${}^2$ Institute for Theoretical Physics, Utrecht University, The
Netherlands. }

\begin{abstract}
The Belinskii, Khalatnikov and Lifshitz conjecture \cite{bkl1}
posits that on approach to a space-like singularity in general
relativity the dynamics are well approximated by `ignoring
spatial derivatives in favor of time derivatives.' In
\cite{ahs1} we examined this idea from within a Hamiltonian
framework and provided a new formulation of the conjecture in
terms of variables well suited to loop quantum gravity. We now
present the details of the analytical part of that
investigation. While our motivation came from quantum
considerations, thanks to some of its new features, our
formulation should be useful also for future analytical and
numerical investigations within general relativity.

\end{abstract}


\maketitle

\section{Introduction}
\label{s1}

Originally formulated in 1970, the Belinskii-Khalatnikov-Lifshitz
(BKL) conjecture states that as one approaches a space-like
singularity, `terms containing time derivatives in Einstein's
equations dominate over those containing spatial derivatives'
\cite{bkl1}. This implies that Einstein's partial differential
equations are well approximated by ordinary differential equations
(ODEs), whence the dynamics of general relativity effectively become
local and oscillatory. The time evolution of fields at each spatial
point is well approximated by that in homogeneous cosmologies,
classified by Bianchi \cite{Bianchi}. The simplest of these are the
Bianchi I metrics which have no spatial curvature and the Bianchi II
metrics which have `minimal' spatial curvature. According to the BKL
conjecture, the dynamics of each spatial point follow the
`Mixmaster' behavior ---a sequence of Bianchi I solutions bridged by
Bianchi II transitions. Finally, with the significant exception of a
scalar field, matter contributions become negligible ---to quote
Wheeler, ``matter doesn't matter".

In the beginning, the conjecture seemed to be coordinate dependent
and rather implausible. However, subsequent analysis by a large
number of authors has shown that it can be made precise and by now
there is an impressive body of numerical and analytical evidence in
its support \cite{Berger}. It is fair to say that we are still quite
far from a proof of the conjecture in the full theory. But there has
been outstanding progress in simpler models. In particular, Berger,
Garfinkle,  Moncrief, Isenberg, Weaver and others showed that, in a
class of models, as the singularity is approached the solutions to
the full Einstein field equations approach the `Velocity Term
Dominated' (VTD) ones obtained by neglecting spatial derivatives
\cite{Berger,Garf,Moncrief,Isenberg,Weaver}. Andersson and Rendall
\cite{ar} showed that for gravity coupled to a massless scalar field
or a stiff fluid, for every solution to the VTD equations there
exists a solution to the full field equations that converges to the
VTD solution as the singularity is approached, \emph{even in the
absence of symmetries}. These results were generalized to also
include $p$-form gauge fields in \cite{dhrw}. In these VTD models
the dynamics are simpler, allowing a precise statement of the
conjecture that could be proven. In the general case, the strongest
evidence to date comes from numerical evolutions. Berger and
Moncrief began a program to analyze generic cosmological
singularities \cite{Berger2}. While the initial work focused on
symmetry reduced cases \cite{Berger3}, more recently Garfinkle
\cite{dg1} has performed numerical evolution of space-times with no
symmetries in which, again, the Mixmaster behavior is apparent.
Finally, additional support for the conjecture has come from a
numerical study of the behavior of test fields near the singularity
of a Schwarzschild black hole \cite{sag}.

With growing evidence for the BKL conjecture, it is natural to
consider its implications to quantum gravity. The conjecture
predicts a dramatic simplification of general relativity near
space-like singularities, which are precisely the places where
quantum gravity effects are expected to dominate. A promising
approach to analyze this issue is provided by loop quantum
cosmology (LQC) \cite{LQC} where there are now several
indications that the quantum gravity effects become important
only when curvature or matter density are about a percent of
the Planck scale. Therefore it is quite possible that,
generically, spatial derivatives become negligible compared to
the time derivatives already when the universe is sufficiently
classical. In this case a quantization of the effective theory
with ODEs, that descends from techniques applicable in the full
theory, could provide a reliable qualitative picture of quantum
gravity effects near generic space-like singularities. If, on
the other hand, the BKL behavior sets in only in the Planck
regime, this strategy would not be viable. But since there is
no reason to trust Einstein's equations in this regime, then
the conjecture would also not have a physically interesting
domain of validity.

LQC is the result of application of the principles of loop quantum
gravity (LQG) \cite{alrev,crbook,ttbook} to symmetry reduced
cosmological models. Initial study of the k=0 Friedmann Lemaitre
Robertson Walker (FLRW) models revealed that the quantum geometry
effects underlying LQG provide a natural mechanism for the
resolution of the big bang singularity \cite{mb1}. Subsequent more
complete analysis led to a detailed understanding of the physics in
the Planck regime and also showed that although these effects are
very strong there ---capable of replacing the big bang with a
quantum bounce--- they die extremely rapidly so as to recover
general relativity as soon as the curvature falls below Planck scale
\cite{aps1}. These results were then extended to include spatial
curvature in \cite{apsv} and a cosmological constant in \cite{bp}.
More recent investigations reveal that if matter satisfies a
non-dissipative equation of state $P = P(\rho)$, LQC resolves all
strong curvature singularities of the FLRW models, including, e.g.,
those of the `big-rip' or `sudden death' type \cite{ps}. Also, it is
now known in LQC that the Bianchi I and II and IX singularities are
resolved \cite{EdB1,EdB2,EdB9}.

In view of the BKL conjecture, these results, together with further
support from the `hybrid' quantization of Gowdy models \cite{Gowdy},
suggest that there may well be a general theorem to the effect that
all space-like singularities of the classical theory are naturally
resolved in LQG. However, it is difficult to test this idea using
the current formulations of the BKL conjecture since these
approaches are motivated by the theory of partial differential
equations rather than by Hamiltonian or quantum considerations (See
e.g., \cite{UEWE,ddb}. In particular, most approaches perform a
rescaling of their dynamical variables by dividing by the trace of
the extrinsic curvature. It is difficult to promote the resulting
variables to operators on the LQC Hilbert space. In the analysis
presented here, we reformulate the BKL conjecture in a way better
suited to LQC and explore the resulting system both analytically and
numerically.

In LQG one begins with a first order formalism where the basic
canonical variables are a density-weighted triad and a
spin-connection \cite{alrev,crbook,ttbook}. In section \ref{s2} we
will begin by recalling this Hamiltonian formulation of general
relativity. In section \ref{s3} we rewrite this theory using a set
of variables that are motivated by the BKL conjecture. Rather
surprisingly, the core of this theory can be formulated using
(density weighted) fields with only internal indices;
\emph{space-time tensors never feature}! To understand the
implications of the BKL conjecture to LQG, we need to express the
conjecture using this Hamiltonian framework. This task is carried
out in section \ref{s4}. We provide a weak and a strong version of
the conjecture. The key idea is to say that, as one approaches
space-like singularities, the exact system is well approximated by a
truncated system which features only time derivatives.
Non-triviality of the formulation lies in the choice of variables
and specification of how limits are taken. Our procedure satisfies a
number of stringent requirements. In particular, one can either
first truncate the Hamiltonian and then obtain the equations of
motion or first obtain the full equations of motion and then
truncate them; the two procedures commute. In section \ref{s5} we
study the truncated Hamiltonian system and explore its dynamics in
some detail. We  show that it exhibits all the known features such
as the `u-map' and spikes. Thus, \emph{the Hamiltonian framework we
were led to by LQG considerations successfully captures the
Mixmaster dynamics faithfully.} Therefore, in addition to providing
a viable point of departure to analyzing the fate of generic
space-like singularities in LQG, it should also be useful in
analytical and numerical investigations of the BKL conjecture in
classical general relativity itself. In section \ref{s6} we
summarize the main results and comment on their relation to those of
other works.

The two appendices contain more technical material. Appendix
\ref{a1} introduces densities in a coordinate-free manner. This
notion is important because the basic variables in our formulation
of the BKL conjectures are scalar densities of weight 1. In the main
text, for simplicity we have set the shift and the Lagrange
multiplier of the Gauss constraint equal to zero. Appendix \ref{a2}
contains the full equations without these restrictions.

\section{Preliminaries}
\label{s2}

We will consider space-times of the form ${}^4\!M = \mathbb{R}
\times {}^3\!M$ where ${}^3\!M$ is a compact, oriented
3-dimensional manifold (without boundary).%
\footnote{The restriction on topology is made primarily to avoid
having to specify boundary conditions and having to keep track of
surface terms. There is no conceptual obstruction to removing this
restriction (following, for example, the Hamiltonian framework
underlying LQC).}
We will formulate general relativity in terms of first order
variables, the point of departure of LQG \cite{aa}. These consist of
pairs of fields consisting of a (density weighted) orthonormal
triad, $\E^a_i$ and its conjugate momentum $K_a^i$ which on
solutions will correspond to extrinsic curvature. The fundamental
poisson bracket is given by
\be
\{ \E^a_i(x), K_b^j(y)\} = \delta_i^j \delta^a_b \delta^3(x-y)
\ee
Herein, early letters, $a,b,c$ denote spatial indices while
$i,j,k$ denote internal indices which take values in $so(3)$
---the Lie algebra of $SO(3)$. Tildes are used to capture density
weights of quantities; a tilde above indicates that the quantity
transforms as a (tensor) density of weight 1 and a tilde below will
denote a (tensor) density of weight $-1$. The internal indices can
be freely raised and lowered using a fixed kinematical metric
$\q_{ij}$ on $so(3)$. The phase space spanned by smooth pairs
$(\E^a_i, K_a^i)$ will be denoted by $\mathcal{P}$.

These phase space variables are related to their Arnowitt, Deser and
Misner (ADM) \cite{ADM} counterparts by
\ba
\E^a_i \E^{b}_j\, \q^{ij}  = \tilde{\tilde{q}}\, q^{ab} \\
K_a^i \E^b_i = \widetilde{\sqrt{q}}\, K_a^{\;\;b} \ea
where $q_{ab}$ is the metric on the leaf ${}^3\!M$, $q$ its
determinant, and $K_{ab}$ the extrinsic curvature of ${}^3\!M$. In
terms of these variables we perform a 3+1 decomposition of
space-time to obtain as Hamiltonian a sum of constraints with
Lagrange multipliers \cite{aa,jr}:
\be H[\tilde{E},K]=\int_{{}^3\!M}- \f{1}{2}\utw{N}\,
\tilde{\tilde{S}} - \f{1}{2} N^a \tilde{V}_a + \Lambda_i
\tilde{G}^i\, . \ee %%
The Lagrange multipliers $\N$, $N^a$, the lapse and shift, are
related to the choice of slicing and time in the standard fashion,
and $\Lambda_i$ is related to rotations in the internal space. Phase
space functions $\tilde{\tilde{S}}$, $\tilde{V}_a$, and
$\tilde{G}^k$ are the scalar, vector, and Gauss constraints (with
density weights $2, 1, 1$ respectively), given by \cite{aa,jr}
\ba \label{Co} &\tilde{\tilde{S}}(\E,K) &\equiv
-\tilde{\tilde{q}}\,\mathcal{R} - 2\E^a_{[i} \E^b_{j]}\,
K_a^i K_b^j \\
&\tilde{V}_a(\E,K) &\equiv 4 D_{[a}(K_{b]}^i\, \E^b_i ) \\
&\tilde{G}^k(\E,K) & \equiv \ep_i^{\; jk} \E^a_j\, K_a^i \ea
Where $\mathcal{R}$ is the scalar curvature of the metric
$q_{ab}$.  The overall sign and numerical factors in the
constraints are chosen so they reduce to the standard ADM
constraints upon solving the Gauss constraint. $\mathcal{R}$
can be written in terms of the triad and its inverse or in
terms of the triad and the connection $\Gamma_a^i$ compatible
with the triad, which is defined by
\be D_a \E^b_i + \ep_{ijk} \Ga_a^j \E^{bk} = 0, \qquad {\rm
or}\quad \Gamma_a^j = -\f{1}{2} \utw{E}_{bk}\, D_a \E^b_i\,
\ep^{ijk}\, . \ee
(Note that $D_a$ acts only on tensor indices; \emph{it treats
the internal indices as scalars.}) Although $\Gamma_a^i$ is
determined entirely by $\E^a_i$ for now it is convenient to use
all three fields $\Gamma_a^i$, $K_a^i$ and $\E^a_i$ in our
classical analysis:  In our formulation of the BKL conjecture
$\Gamma_a^i$ and $K_a^i$ will be the relevant degrees of
freedom near the singularity, so it is natural to express the
theory in terms of them.

The equations of motion are obtained by taking Poisson brackets with
the Hamiltonian on the phase space $\mathcal{P}$:
\ba
\label{Edot0}\dot{\E}^a_i &=& \{\E^a_i,H[\E,K]\} \\
\label{Kdot0}\dot{K}_a^i &=& \{ K_a^i,H[\E,K] \} \, .\ea
$\mathcal{P}$ is the phase space underlying LQG. The basic variables
$(A_a^i, E^a_i)$ used there are obtained by a simple canonical
transformation on $\mathcal{P}$ \cite{aa}:
\be \big(\E^a_i,K_a^i \big) \rightarrow \big(A_a^i,\gamma^{-1}
\E_a^i\big) \quad {\rm with} \quad  A_a^i = \Ga_a^i + \gamma K_a^i\,
, \ee
$\gamma$ being the Barbero-Immirzi parameter of LQC. (In classical
general relativity, space-time equations of motion are independent
of the value of this real parameter.) For simplicity of presentation
we will introduce our formulation of the BKL conjecture using
$(\E^a_i,K_a^i)$ although it will be clear that our framework can be
readily recast in terms of $(\E^a_i, A_a^i)$.

\section{Variables motivated by the BKL conjecture}
\label{s3}

In order to formulate the BKL conjecture in this system, one needs
to specify two things:  What kind of derivatives are to dominate as
one approaches the singularity and what kind are to become
negligible? And, what are the quantities whose derivatives are to be
treated as negligible?  In this section we first motivate and
introduce a set of variables and a derivative operator and then use
them to formulate the conjecture. The main idea is as follows. The
accumulated evidence to date suggests that the spatial metric
$q_{ab}$ becomes degenerate at the space-like singularity whence its
determinant $q$ vanishes there. (In particular, this is borne out in
the numerical simulations of solutions with two commuting Killing
fields ---the so-called $G2$ space-times which include Gowdy models
\cite{WCL1}.) We will focus on the class of singularities where this
occurs. In this case one would expect that if we rescaled fields
which are ordinarily divergent at the singularity with appropriate
powers of $q$, the rescaled quantities would have well defined
limits.

Now, the density weighted triad $\E^a_i$ is obtained by rescaling of
the orthonormal triad $e^a_i$, which is divergent at the
singularity, by $\sqrt{q}$. In examples, the factor of $\sqrt{q}$
not only gives $\E^a_i$ a well defined limit, but the limit in fact
vanishes. Therefore, contraction by $\E^a_i$ can serve to tame
fields which would otherwise have been divergent at the singularity.
This consideration leads us to construct scalar densities by
contracting $\E^a_i$ with $K_a^i$, and $\Ga_a^i$. As noted above,
since contraction with $\E^a_i$ will suppresses the divergence of
$K_a^i$ and $\Ga_a^i$, the combination is expected to remain finite
at the singularity. Let us then set
\ba \label{CP}
\Kbart_i^{\;\,j} &:= \E^a_i K_a^j - \E^a_k K_a^k \delta_i^{\;\,j} \\
\Ct_i^{\;\,j} &:= \E^a_i \Ga_a^j - \E^a_k \Ga_a^k
\delta_i^{\;\,j}\, . \ea
These two fields, $\Kbart_i{}^j$ and $\Ct_i{}^j$ will turn out to be
the relevant variables near the singularity in our BKL framework. In
particular, we will show below that the constraints of general
relativity can be expressed in terms of polynomials of these basic
variables and their derivatives. Therefore if the basic variables
and their derivatives remain finite at the singularity, the
constraints will also continue to hold there. Since the Hamiltonian
of the theory is a linear combination of these constraints, dynamics
of the basic variables will meaningfully extend to the singularity.

Beyond the possibility of being bounded at the singularity, an
important feature of these variables is that they have \emph{only
internal indices} which can be freely raised and lowered using the
fixed, kinematic, internal metric $\q^{ij}$; the dynamical metric
$q^{ab}$ which diverges at singularities is not needed. Under
diffeomorphisms $\Kbart_i{}^j$ and $\Ct_i{}^j$ transform as density
weighted \emph{scalars} on ${}^3\!M$. Because of this feature,
statements about their asymptotic properties can be formulated much
more easily than would be possible if they were tensor fields. (For
a coordinate free introduction to densities, see Appendix \ref{a1}).
% easily. Had they been tensor
%fields, we would have the complication of finding suitable frames to
%evaluate their components in before taking limits.

To illustrate why these variables are likely to be well defined at
the singularity, let us consider the Bianchi I model. Because of
spatial flatness, we can work in an internal gauge in which
$C_i{}^j=0$ everywhere. What about $\E^a_i$ and $P_i{}^j$? In terms
of the commonly used proper time $\tau$, the metric is given by
$ds^2 = -d\tau^2 + \sum_i \tau^{2p_i}dx_i^2$ and the singularity
occurs at $\tau=0$. Since $\sum p_i=1$, we have $q =\tau^2$ in the
Bianchi I chart. In addition, due to the second constraint on the
exponents, $\sum p_i^2 = 1$ whence the density weighted triad
$\E^a_i$ vanishes at the singularity as  $\tau^{1-p_i}$ and
$\Kbar_i^j$ is finite there for each $i$.

We further introduce a derivative operator $\Dt_i$ defined by the
contraction of $D_a$ with $\E^a_i$:
\be \Dt_i :=\E^a_i D_a \, . \ee
The expectation is that this contraction will have the effect of
suppressing terms containing $\Dt_i$ as we approach the singularity.
Thus, $\Dt_i$ will be the spatial derivatives we were seeking which,
when acting on certain quantities, will be conjectured to be
negligible near the singularity.%
\footnote{This operator is linear and satisfies the Leibnitz rule.
It ignores internal indices (since the action of $D_a$ is
non-trivial only on tensor indices). However since its action on a
function $f$ does not yield the exterior derivative $df$, $\Dt_i$ is
not a connection. If we were to formally treat as a connection, it
would have torsion, which is related to $\C$: $\Dt_{[i}\Dt_{j]} f =
-\tilde{T}^k_{\;\;ij} \Dt_k f $ where $\tilde{T}^k_{\;\;ij} =
\ep_{kl[i} \Ct_{j]}^{\;\;l}$. In what follows, $D_i$ often acts on
scalar densities. This action is given explicitly in Appendix
\ref{a1}.}
The variable $\Pt_{ij}$ is related to the momentum
$\Pt^{ab}$(conjugate to the 3-metric) in the ADM phase space by
$\tilde{\tilde{q}}\, \Pt^{ab} = \E^a_i \E^b_j \Pt^{ij}$. $\Ct_{ij}$
encodes information in the $\Dt_i$ spatial derivatives of the triad
$\E^a_i$:
\be \label{Ga} {\Ct}^{ij} = - \utw{E}_a^i\, \ep^{klj}\, {\Dt}_k
\E^a_l \, . \ee
Note that, although the $\Ct_{ij}$ depend on spatial derivatives of
the triad and are often sub-dominant to $\Pt_{ij}$, it turns out
that they are not always negligible in the approach to the
singularity. Indeed, this behavior is observed in the truncated
system, which is discussed in section \ref{s4}. It is $\Ct^{ij}$
rather than the triads themselves that will feature directly in our
formulation of the conjecture.

For simplicity of notation, \emph{from now on we will drop the
tildes.} Thus, from now on each of $E^a_i, \C_i{}^j, \P_i{}^j, \D_i$
carries  a density weight $1$, while the lapse field $N$ carries a
density weight $-1$. The scalar and the vector constraint functions
$S$ and $V_i$ (introduce below) carry density weight 2 while the
Gauss constraint $G^k$ carries density weight 1.

By making use of (\ref{CP}) and (\ref{Ga}), functions of
$(E^a_i,K_a^i)$ and their covariant derivatives can be rewritten in
terms of $(E^a_i, \C_i{}^j, \P_i{}^j)$ and their $\D_i$ derivatives.
The scalar curvature $\mathcal{R}$ for example can be expressed
entirely in terms of $\C_i{}^j$ and its $\D_i$ derivatives:
\be
q\mathcal{R}=-2\ep^{ijk} \D_i (\C_{jk}) - 4 \C_{[ij]} \C^{[ij]} - \C_{ij}
\C^{ji} + \f{1}{2} \C^2
\ee
Consequently, the constraints can be re-expressed
\emph{entirely} in terms of $\C_i{}^j, \P_i{}^j$ and their
$\D_i$ derivatives (with no direct reference to $E^a_i$ or even
the determinant $q$ of the 3-metric):
\ba \label{S1} {S} &=& 2\ep^{ijk} \D_i (\C_{jk}) + 4 \C_{[ij]}
\C^{[ij]} + \C_{ij} \C^{ji} - \f{1}{2} \C^2 +
\P_{ij} \P^{ji} - \f{1}{2} \P^2 \\
\label{V1} {V}_i &=& - 2\D_j \P_{i}{}^j - 2 \epsilon_{jkl}
\P_i{}^j\, \C^{kl} - \epsilon_{ijk} \C \P^{jk} + 2\ep_{ijk} \P^{jl} \C_l{}^k\\
\label{G1} {G}^{k} &=& \epsilon^{ijk} \P_{ji}  \, .\ea
Here we have converted the co-vector index on the vector constraint
${V}_a$ to an internal index by contracting it with ${E}^a_i$. Since
the $E^a_i$ is assumed to be non-degenerate away from the
singularity the constraint ${V}_i$ defines the same constraint
surface as the original vector constraint introduced in (\ref{Co}).
Notice here that the constraint can be easily decomposed into those
terms that contain the derivative $\D_i$ and those that don't.

The equations of motion for $E^a_i$, $\C_i{}^j$, $\Kbar_i{}^j$
can be written in a similar form.  These can be obtained using
the full Poisson brackets (\ref{Edot0}),(\ref{Kdot0}) or by
directly computing Poisson brackets of $\Kbar_i{}^j$ and
$\C_i{}^j$ with the scalar/Hamiltonian constraint. To
streamline the second calculation, let us specify the Poisson
brackets between $E^a_i$, $\C_i{}^j$, and $\Kbar_i{}^j$:
\ba \label{pb1} \{{E}^a_i(x), \Kbar_j{}^k(y) \} &=& \big({E}^a_j(x)
\delta_i{}^k- {E}^a_i (x)\delta_j{}^k\big)\, \delta(x,y) \\
\label{pb2} \{ \Kbar_i{}^j(x),\Kbar_k{}^l(y) \} &=&
\big(\Kbar_k{}^j(x) \delta_i{}^l - \Kbar_i{}^l(x) \delta_k{}^j\big)
 \delta(x,y) \\
\label{pb3} \{ {\textstyle{\int}} f_{ij} \Kbar^{ij},
{\textstyle{\int}} g_{kl} \C^{kl} \} &=& {\textstyle{\int}}
\big( f_{ij} g_{kl} (\C^{kj} \delta^{il} + \C^{jl} \delta^{ik})
+\ep^{jlm} \delta^{ik} g_{kl} \D_m f_{ij}\big)\\
\label{pb4} \{E^a_i(x), \C^j{}_k(y)\} &=& 0 \quad {\rm and}
\quad \{\C_i{}^j(x),\C_k{}^l(y)\} =0\, , \ea
where $f_{ij}, g_{ij}$ are smooth test scalar fields. The equations
of motion obtained by taking Poisson brackets with the scalar
constraint are then given by
\ba \label{Cdot1}&\dot{\C}^{ij} = - &\ep^{jkl}  \D_k ({N}
(\f{1}{2} \delta_l^i \Kbar-\Kbar_l^{\;\,i})) + {N} [2
\C^{(i}_{\;\;k} \Kbar^{|k|j)} +
2\C^{[kj]} \Kbar_k^{\;\,i} - \Kbar \C^{ij}]\\
\label{Pdot1}&\dot{\Kbar}^{ij} = &\ep^{jkl}  \D_k ({N} (1/2
\delta_l^i \C-\C_l^{\;\,i}))  - \ep^{klm} \D_m ({N} \C_{kl})
\delta^{ij}+
2\ep^{jkm} \C^{[ik]} \D_m ({N})\nonumber \\
& & (\D^i \D^j - \D^k \D_k \delta^{ij}) {N} + {N} [- 2
\C^{(ik)} \C_k^{\;\,j} + \C \C^{ij}- 2 \C^{[kl]}
\C_{[kl]}\delta^{ij}] \ea
and
$$\dot{E^a_i} =  -{N} \Kbar_i^{\;\,j} E^a_j$$
where we have set the shift to zero to reduce clutter. (For non-zero
shift, see Appendix \ref{a2}.) Note that the equation of motion for
$E^a_i$ is a simple ODE. Note also that, as was the case with
constraints, the equations of motion for $\C_i{}^j$ and
$\Kbar_i{}^j$ can again be written in terms of scalar densities and
the derivative $\D_i$ \emph{only}. This motivates us to ask for an
evolution equation for the derivative operator $D_i$. Since $D_i$
ignores internal indices, it suffices to consider its action just on
scalar densities $S_n$ of weight $n$. We have:
\be \label{Ddot1} \dot{\D}_i S_n = \f{n}{2}\, [\D_i (N \P)]S_n
\,-\, N \P_i{}^j\D_j S_n\, . \ee

\emph{Thus we have cast all the constraint as well as evolution
equations as a closed system involving only} $\C_i{}^j$,
$\Kbar_i{}^j$, and $\D_i$. These equations can then be used as
follows. On an initial slice, we construct $(\C_i{}^j, \P_i{}^j,
\D_i)$ from a pair $(E^a_i, K_a^i)$ of canonical variables. Then we
can deal exclusively with the triplet $(\C_i{}^j, \P_i{}^j, \D_i)$.
The pair $(E^a_i, K_a^i)$ satisfies constraints if and only if the
triplet satisfies (\ref{S1})--(\ref{G1}). Given such a triplet, we
can evolve it using (\ref{Cdot1}),(\ref{Pdot1}),(\ref{Ddot1})
\emph{without having to refer back to the original canonical pair}
$(E^a_i, K_a^i)$. These two sets of equations have some interesting
unforeseen features. First, as already mentioned, the basic triplet
$(C_i{}^j, P_i{}^j, D_i)$ has \emph{only internal indices}: our
basic fields are \emph{scalars} on ${}^3\!M$ (with density weight
1). It would be of considerable interest to investigate if this fact
provides new insights into the dynamics of 3+1 dimensional gravity
\cite{bh}. Second, these equations \emph{do not refer to the triad}
$E^a_i$. Suppose we begin at an initial time where $C_i{}^j$ is
derived from an $E^a_i$. Then these constraint and evolution
equations ensure that $C_i{}^j$ \emph{is derivable from a triad at
all times}. Furthermore, we can easily construct that triad directly
from a solution $(C_i{}^j, P_i{}^j)$ to these equations: first solve
(\ref{Cdot1})--(\ref{Ddot1}) and then simply integrate the ODE
\be \label{Edot} \dot{E}^a_i =  -{N} P_i{}^j E^a_j\,  \ee
at the end. Third, the structure of the constraint and evolution
equations in terms of $(C_i{}^j, P_i{}^j, D_i)$ is remarkably simple
since only low order polynomials of these variables are involved.
Finally, thanks to our rescaling by $\sqrt{q}$, our basic triplet
$C_i{}^j, P_i{}^j, D_i$ (as well as $E^a_i$) are expected to have a
well behaved limit at the singularity. A close examination of our
equations shows that they allow the triad become to become
degenerate during evolution. So, strictly (as in LQG \cite{aa,jr})
we have a generalization of Einstein's equations.

To summarize, we have found variables which remain finite at the
singularity in examples and rewritten Einstein's equations as a
\emph{closed system of differential equations} in terms of them.
Therefore, this formulation may be useful for proving global
existence and uniqueness results and rigorous exploration of fields
near space-like singularities. Finally, although for simplicity we
have set shift $N^i$ and the smearing field $\Lambda_i$ equal to
zero, the features we just discussed hold more generally (see
Appendix \ref{a2}).

To conclude, let us examine the action of the vector and the
Gauss constraints on our basic variables. (The action of the
scalar constraint yields the evolution equations which we have
already discussed.) Since the vector constraint generates a
combination of spatial diffeomorphisms and internal rotations,
it is standard to subtract a multiple of the Gauss constraint
to define the diffeomorphism constraint:
\be \label{Diff} V'_i = V_i - 2(\C_i{}^j - \frac{\C}{2}
\delta_i{}^j) \, G_j \ee
We can then smear both constraints to obtain
\ba G[\Lambda] = \int_{{}^3\!M} \Lambda^k G_k \quad {\rm and} \quad
 V'[N] =\int_{{}^3\!M} N^i V'_i \ea
where $N^i$ s a scalar with density weight $-1$ so that $N^a:=
N^iE^a_i$ is the standard lapse and, as before, $\Lambda^i$ has
density weight zero. The action of $G[\Lambda]$ on the basic
variables is given as usual via Poisson brackets:
\ba \label{gt1}\{P_{ij},G[\Lambda] \} &=& \ep_{klj} \Lambda^l
P_i{}^k +
\ep_{kli} \Lambda^l P^k{}_j \\
\label{gt2} \{ C_{ij},G[\Lambda] \} &=& \ep_{klj} \Lambda^l
C_i{}^k +
\ep_{kli} \Lambda^l C^k{}_j + D_i \Lambda_j - D_k \Lambda^k \delta_{ij}\\
\label{gt3}\{ D_i S_n, G[\Lambda] \} &=& \ep_{jki} \Lambda^k
D^j S_n \ea
In the last equation $S_n$ is any scalar density of weight n. As
expected the Gauss constraint generates infinitesimal $SO(3)$
transformations with $D_i S_n$ and $P_i{}^j$ transforming as tensors
and $C_i{}^j$ transforming as (the contraction of a triad with) an
$SO(3)$ connection.

Similarly, the action of the diffeomorphism constraint is given
by the Poisson brackets:
\ba \label{vt1}\{ P_i{}^j, V'[N] \} &=& -2(N^k D_k P_i{}^j + P_i{}^j
D_k N^k + P_i{}^j \ep_{klm} N^k C^{lm}) = -2\mathcal{L}_{\vec{N}} P_i{}^j\\
\label{vt2} \{ C_i{}^j, V'[N] \} &=& -2(N^k D_k C_i{}^j + C_i{}^j
D_k N^k + C_i{}^j \ep_{klm} N^k C^{lm}) = -2\mathcal{L}_{\vec{N}} C_i{}^j\\
\label{vt3} \{ D_i S_n, V'[N] \} &=& -2(N^j D_j (D_i s_n) + n
D_j (N^j) D_i S_n - n \ep_{jkl} N^j C^{kl} D_i S_n) = -2
\mathcal{L}_{\vec{ N}} D_i S_n\nonumber\\ \ea
where $\vec{N} \equiv N^a = E^a_i N_i$. We see that the
constraint generates diffeomorphisms as expected with
$P_i{}^j$,\,$C_i{}^j$, and $D_iS_n$ transforming as scalar
densities. Again, note that the infinitesimal changes generated
by each constraint involve \emph{only} the basic variables
$\C_i{}^j$, $\P_i{}^j$, and $\D_i$. Thus there is still a
closed system in terms of this set of variables.

\section{The conjecture}
\label{s4}

In order to express the BKL conjecture we must make more precise the
arena in which it is to be applied. The ingredients we need are a
space-time with a space-like singularity, a notion of `spatial' and
`temporal' derivatives, and specification of the system to which the
conjecture is to be applied. We make use of the framework introduced
in the previous section to provide this arena.

Let us begin with a 4-manifold, ${}^4\!M$ admitting a smooth
foliation $M_t$ parameterized by a time function, $t$. We restrict
ourselves to a slicing of ${}^4\!M$ in which the space-like
singularity lies on the limiting leaf. This ensures that we can
reasonably discuss an approach to the singularity as approaching the
limiting leaf. The time function $t$ labeling our spatial slices is
intertwined with the choice of lapse and shift. We will assume that
the lapse $N$ and the shift $N^i$, \emph{each with density weight}
$-1$, admit a smooth limit as one approaches the singularity. Since
the spatial metric $q_{ab}(t)$ becomes degenerate at the
singularity, the commonly used lapse function $\bar{N}:= \sqrt{q}N$
(with density weight zero) goes to zero, thus placing the
singularity at $t=\infty$. (These assumptions are minimal and
further constraints on admissible foliations may well be needed in a
more complete framework.)

Our basic variables will be $(C_i{}^j, P_i{}^j)$, the lapse $N$, and
the shift $N^i$. By \emph{time derivatives}, we will mean their Lie
derivatives along the vector field $t^a := \bar{N} n^a+N^a$ where
$n^a$ is the unit normal to the foliation $M_t$. By \emph{spatial
derivatives} we will mean their $D_i$ derivatives. Since $D_i :=
E^a_i D_a$, the notion does not depend on coordinates. Rather, it is
tied directly to the physical triads and the covariant derivatives
compatible with them. Then, the idea behind the conjecture is that,
\emph{as one approaches the singularity, the spatial derivatives
$D_i C_j{}^k,\, D_iP_j{}^k, \, D_i N, \, D_i N^j$ of the basic
fields should become negligible compared to the basic fields
themselves} because of the $\sqrt{q}$ multiplier in the definition
of $E^a_i$ which descends to $D_i$.

We now show that an immediate consequence of this assumption is
that the antisymmetric part of $C_{ij}$ is negligible compared
to the other basic fields. Let us define $a^i := \ep^{ijk}
C_{jk}$. Then by conjecture $D_i (a^i N)$ is negligible.%
\footnote{By their definitions, the internal metric $\q_{ij}$
and the alternating tensor $\epsilon_{ijk}$ are kinematic,
fixed once and for all, and are annihilated by all derivative
operators $D_a$ and $D_i$.}
Since the spatial manifold is assumed to be compact, integrating
this negligible quantity and then integrating by parts, one obtains:
\be \int_{{}^3\!M} D_i (N a^i) = \int_{{}^3\!M} N a^i\, D_a {E}^a_i
= \int_{{}^3\!M} N\, a^i a_i\, ,\ee
where we have used the definition of $C_{ij}$ which implies $D_a
{E^a_i} = \epsilon_{ijk} C^{jk}$. Since the internal metric and the
lapse are positive, we conclude that $a_i$ and hence $C_{[ij]}$ are
necessarily negligible under our assumptions. This fact will be
useful throughout our analysis.

Next, note that we have expressed general relativity in the form of
a constrained theory in terms of our basic variables,
$C_i{}^j$,$P_i{}^j$, and $D_i$. Our constraints are composed of
quadratic terms in our basic variables and terms of the form $D_i
C_j{}^k$ and $D_i P_j{}^k$. We can therefore split each constraint
into two parts ---terms which contain no derivatives and those which
do. Similarly the equations of motion can be split into terms that
contain derivatives and those that do not. With this background, we
can state two versions of our conjecture.\newline

\bf Weak Conjecture : \rm As the singularity is approached the terms
containing derivatives in the constraints and equations of motion
are negligible in comparison to the polynomial terms. Thus, as the
singularity is approached the constraints and equations of motion
approach those found by setting derivative terms to zero. \newline

We define the truncated theory to be the system defined by setting
$D_i$-derivative terms to zero,
\be \label{truncdef}\D_i \C_j{}^k=\D_i \Kbar_j{}^k=\D_i N=\D_i N_j
=\C_{[ij]}=0\, , \ee
in the equations of motion (\ref{Cdot1})-(\ref{Ddot1}) and
constraints (\ref{S1})-(\ref{G1}). Thus, the weak conjecture says
that the equations of motion can be well approximated by those of
the truncated theory in the vicinity of the singularity.  Note that
this does not imply that the \emph{solutions} of the full equations
of motion will approach the solutions to the truncated equations as
the singularity is approached.  This additional condition is
captured in the strong version as follows.
\newline

\bf Strong Conjecture: \rm As the singularity is approached the
constraints and the equations of motion approach those of the
truncated theory and in addition the solutions to the full
equations are well approximated by solutions to the truncated
equations. \newline

With the strong conjecture the solution of the full Einstein
equations will asymptote to solutions of the truncated system
defined by (\ref{truncdef}).  In the following we will analyze this
truncated system.

Not only are the truncated constraints purely algebraic, but
they involve only quadratic combinations our basic variables:
\ba \label{S2} S_{(T)} &:=&  \C_{ij} \C^{ji} - \f{1}{2} \C^2 +
\P_{ij} \P^{ji} - \f{1}{2} \P^2 \\
\label{V2} V_i^{(T)} &:=&  - \ep_{ijk}\, \C \P^{jk} + 2\ep_{ijk}
\P^{jl} \C_l^{\;\,k} \\
\label{G2} G^{k}_{(T)} &:=& \epsilon^{ijk} \P_{ji} ,
\ea
The truncated Gauss constraint is in fact exact because
(\ref{G1}) involves no derivative terms, while the scalar and
the diffeomorphism constraints are genuinely truncated.

The infinitesimal transformations (\ref{gt1}) - (\ref{vt3})
generated by the full constraints contain derivative terms that
are now assumed to be negligible in comparison to the
polynomial terms. Ignoring the negligible terms leads us to the
following transformations on the basic fields:
\ba \label{GT1} \{P_{ij},\, G(\Lambda)\}_{T} &=&
2 \ep_{kl(j} P_{i)}{}^{k} \Lambda^l \\
\label{GT2} \{C_{kl},\, G(\Lambda)\}_{T} &=&
2 \ep_{kl(j} C_{i)}{}^{k} \Lambda^l\\
\label{VT1} \{P_{ij},\,  V(N)\}_{T} &=&  4 \ep_{kl(j}
P_{i)}{}^{k} N^m\, (C_m{}^l - \frac{C}{2} \delta_m{}^l)\\
\label{VT2} \{C_{ij},\,  V(N)\}_{T} &=& 4 \ep_{kl(j}
C_{i)}{}^{k} N^m\, (C_m{}^l - \frac{C}{2} \delta_m{}^l) \\
\label{ST1} \{{C}_{ij}, S(N)\}_{T} &=& -2N\, (2 C_{k(i} P^k{}_{j)} - PC_{ij}) \\
\label{ST2} \{{P}_{ij},S(N)\}_{T} &=& -2N\, (- 2 C_{ik} C^k{}_j + C
C_{ij})\,  . \ea

The Gauss constraint continues to generate internal rotations, but,
whereas in the full theory $C_i{}^j$ transforms as (the contraction
of the triad with) a connection,  after truncation both $C_i{}^j$
and $P_i{}^j$ transform as $SO(3)$ tensors.  The vector constraint
also generates internal rotations, since the diffeomorphism
constraint generates only negligible terms.

We arrived at the truncated equations of motion by first obtaining
the full equations and then applying the truncation to them i.e., by
setting spatial derivative terms to zero. But we could also have
first truncated the constraints to obtain (\ref{S2})- (\ref{G2}) and
then computed their truncated Poisson brackets with the basic
variables. This leads to a consistency check of our scheme: do the
two procedure yield the same `truncated equations of motion' in the
end? The answer is in the affirmative. This fact is illustrated by
the following `commutativity diagram':

$ \begin{CD}
    \\
    \text{Full Constraint} @> \text{Truncation} >>
    \text{Truncated Constraint} \\
    @VV\text{Equation of Motion}V
    @VV\text{Equation of Motion}V \\
    \text{Full Equation of Motion}
    @>\text{Truncation} >> \text{Truncated Equation of Motion}
    \\
    \\
    \end{CD} $

Note that the operation of truncation, the final truncated system,
and hence the consistency requirement mentioned above depends
crucially on one's choice of basic variables and notions of space
and time derivatives. For example, if we had adopted the more
`obvious' strategy and used triads $E^a_i$ rather than $\C_i{}^j$ as
basic variables, we would have been led to set $\C_i{}^j$ to zero in
the truncation procedure since $\C_i{}^j$ would then be derived
quantities, obtained by taking the $D_i$ derivative of $E^a_i$. This
truncation would have led us just to Bianchi I equations. \emph{The
resulting BKL conjecture would have been manifestly false.} Thus,
considerable care is needed to arrive at variables which satisfy a
closed set of equations in a Hamiltonian framework, suggest a
natural way to make the heuristic idea of ignoring spatial
derivatives in favor of time derivatives precise, lead to the above
commuting diagram, and a version of the BKL conjecture that is
compatible with the large body of analytical and numerical results
that has accumulated so far. It is rather striking that the
variables $(C_i{}^j, P_i{}^j)$ automatically satisfy these rather
stringent criteria.

\section{Hamiltonian formulation of the truncated system}
\label{s5}

In this section we will analyze the truncated system in some detail
and show that its solutions reproduce the expected BKL behavior. The
section is divided into three parts. In the first we regard
$C_i{}^j, P_i{}^j$ as fields on the full phase space $\mathcal{P}$,
obtain the truncated Poisson brackets between them and truncated
constraints. In the second we solve and gauge fix the vector and the
Gauss constraints of the truncated theory. The result is a finite
dimensional, reduced phase space with a single constraint which is
well suited to serve as a starting point for quantization inspired
by the BKL conjecture. In the third part we discuss several features
of solutions to this Hamiltonian theory. In particular, we will find
that they exhibit Bianchi I phases with Bianchi II transitions.

\subsection{Truncated Poisson brackets}
\label{s5.1}

Since the truncated equations of motion can be formulated entirely
in terms of $C_i{}^j, P_i{}^j$, let us truncate the Poisson brackets
(\ref{pb2}), (\ref{pb3}) we obtained between them by setting the
negligible terms on the right side to zero. Since the full Poisson
bracket (\ref{pb3}) involves smearing fields $f_{ij}$ and $g_{ij}$
we first need to specify which terms involving them are to be
regarded as negligible. The most natural avenue is to construct to
$f_{ij}$ and $g_{ij}$ only from the basic fields
$(\C_i{}^j,P_i{}^j,N, N_i,\q_{ij}, \epsilon^{ijk})$ (and their $D_i$
derivatives). Then the terms containing $\D_i$ derivatives of the
smearing fields will also be negligible and hence vanish in the
truncation. The resulting truncated Poisson brackets between
$C_i{}^j$ and $P_i{}^j$ are then given by:
\ba \label{TPB1}    \{\Kbar_i{}^j(x), \C_k{}^l(y) \}_{T} &=&
\big(\C_k{}^j \delta_i{}^l + \C^{jl} \delta_{ik}\big)(x)\,
\delta(x,y) \\
\label{TPB2} \{ \Kbar_i{}^j(x), \Kbar_k{}^l(y)\}_{T} &=&
\big(\Kbar_k{}^j \delta_i{}^l - \Kbar_i{}^l \delta_k{}^j\big)(x)\,
\delta(x,y)  \\
\label{TPB3} \{\C_i{}^j(x), \C_k{}^l(y) \}_{T} &=& 0\, . \ea
These Poisson brackets suffice to determine the equations of motion
because the truncated Hamiltonian constraint (\ref{S2}) is algebraic
in $C_i{}^j$ and $P_i{}^j$. They are now ODEs,
\be \label{Cdot2}\dot{C}_{ij} =   {N}\, [2 C_{k(i} P^k{}_{j)} - P
C_{ij}] \quad {\rm and} \quad \dot{P}_{ij} = N [- 2 C_{ik} C^k{}_j +
C C_{ij}]\, , \ee
so the truncated dynamics at any one spatial point decouple from
those at other points.

This system has some notable features. First, we have a \emph{closed
system} expressed entirely in terms of $\C_i{}^j(x)$ and
$\Kbar_i{}^j(x)$ at any fixed point $x$. Furthermore, the equations
of motion (\ref{Cdot2}) and constraints (\ref{S2})-(\ref{G2}) are at
most quadratic in these variables. In the full theory, the triad
does not appear explicitly in the equations (\ref{Cdot1}) -
(\ref{Ddot1}) but is implicitly present through $\D_i$. Upon
truncation, even this implicit dependence disappears. Second, as in
the full theory, one can first solve the equations of motion for
$\C_i{}^j(x)$ and $\Kbar_i{}^j(x)$ and then evolve the triad at that
point at the end by solving an ODE. Third, the truncated scalar
constraint (\ref{S2}) is symmetric under interchange of $\C_i{}^j$
and $\Kbar_i{}^j$ and, by adding a multiple of the Gauss constraint,
the vector constraint can be made anti-symmetric under this
interchange:
\be \label{vred} {\bar{V}}^i_{(T)} := \ep^{ijk} \Kbar_j{}^l
\C_{kl}\, . \ee
However, this symmetry is broken at the level of equations of motion
because the truncated Poisson algebra does not have a simple
transformation property under this interchange.

Because fields at distinct points decouple, to study the truncated
system from the viewpoint of differential equations, one can simply
restrict oneself to a single spatial point. However, this is not
directly possible in the Hamiltonian framework because even in the
truncated theory, the Poisson brackets (\ref{TPB1}), (\ref{TPB2})
involve $\delta(x,y)$. But one can introduce a \emph{subspace}
$\mathcal{P}_{\rm hom}$ of the full phase space $\mathcal{P}$
tailored to our truncation. Given a point $(E^a_i, K_a^i)$ in
$\mathcal{P}$ consider the pair $(C_i{}^j, P_i{}^j)$ of density
weighted fields it determines. The phase space point will be said to
be \emph{homogeneous} if there exists an internal gauge and a
nowhere vanishing scalar density $S_{-1}$ of weight $-1$ such that
the (density weight zero) scalar fields $(S_{-1}C_i{}^j,
S_{-1}P_i{}^j)$ are constants on ${}^3\!M$ (and $C_{[ij]} =0$).
(Fixing a $S_1$ is equivalent to fixing a 3-form on ${}^3\!M$; see
Appendix \ref{a1}.) Clearly, the truncated dynamics leaves this
\emph{homogeneous sub-space} $\mathcal{P}_{\rm hom}$ of the phase
space invariant. More importantly, $\mathcal{P}_{\rm hom}$ \emph{is
invariant under full dynamics}: If the $D_i$-derivatives are
initially zero they remain zero under the full equations of motion.
The Hamiltonian dynamics on $\mathcal{P}_{\rm hom}$ fully captures
the truncated dynamics at any fixed spatial point on ${}^3\!M$.

\emph{Remark:} Since the triads $E^a_i$ in the full phase space
$\mathcal{P}$ have been assumed to be non-degenerate, they are
also non-degenerate in $\mathcal{P}_{\rm hom}$. However, as
examples suggest, one would expect them to be become degenerate
in the limit to the space-like singularity where, however,
$C_i{}^j, P_i{}^j$ would continue to be well behaved (and some
of them may even vanish). It is therefore of some interest to
extend the homogeneous subspace by adding `limit points' which
have this behavior. This construction is not needed in our
analysis. However, since it may be useful in future
investigations, we will conclude this subsection with a brief
summary. Let us allow the density weighted triads $E^a_i$ to
become degenerate such that the subspaces spanned by the
non-degenerate directions of vector fields $S_{-1}E^a_i$ are
integrable. (If this condition is satisfied for one nowhere
vanishing scalar density $S_{-1}$, it is satisfied for all.)
Thus, in the degenerate case we obtain preferred 2 or 1
dimensional sub-manifolds on ${}^3\!M$. We can extend the phase
space by including such degenerate $E^a_i$ if, in addition, the
resulting pair $(C_i{}^j, P_i{}^j)$ is regular, $C_{ij}$ is
symmetric and the pair $S_{-1} C_i{}^j, S_{-1}P_i{}^j$ is
homogeneous along the preferred lower dimensional sub-manifolds
of ${}^3\!M$. Key questions for the BKL conjecture are then:
i) Does the Hamiltonian flow on $\mathcal{P}$ naturally extend to
this extension?; and ii) Do generic dynamical trajectories flow
to it?

\subsection{Reduced Phase Space}
\label{s5.2}

Since $C_{ij}$ is symmetric but $P_{ij}$ is not, the
homogeneous subspace $\mathcal{P}_{\rm hom}$ is not a
symplectic sub-manifold of the full phase space $\mathcal{P}$.
But it turns out that one can obtain a symplectic manifold by
solving and gauge fixing the truncated vector and the Gauss
constraints. It will be referred to as the \emph{reduced phase
space}, $\mathcal{P}_{\rm red}$.

The Gauss constraint (\ref{G2}) is equivalent to asking that
$P_{ij}$ is symmetric and then the vector constraint (\ref{V2}) is
equivalent to asking that as matrices, $C_i{}^j$ and $P_i{}^j$
should commute. To gauge fix the Gauss constraint, we first note the
transformation properties (\ref{GT1}) and (\ref{GT2}) of $P_i{}^j$
and $C_i{}^j$ under the action of the Gauss constraint. It is easy
to verify that, because $P_i{}^j$ and $C_i{}^j$ commute, the
requirement that they both be diagonal gauge-fixes the Gauss
constraint completely. It turns out that the diagonality requirement
also fixes the vector constraint. This may seem surprising at first.
But note that the combination $\bar{V}$ of the vector and the Gauss
constraint of Eq (\ref{vred}) again generates internal gauge
rotations, where, however the generator $\Lambda^i$ is a `q-number',
i.e., depends on the phase space variables: $\Lambda^i = N^j(C^i{}_j
- C \delta^i_j)$, where $N^j$ is the shift used to smear the vector
constraint. The fact that the gauge fixing of the vector constraint
does not impose additional requirements on $(C_{ij}, P_{ij})$
`cures' the mismatch in the degrees of freedom in the homogeneous
subspace (arising from the fact that while $C_{ij}$ is symmetric,
$P^{ij}$ is not.).

So far $C_i{}^j, P_i^j$ are fields on ${}^3\!M$, each carrying
density weight 1. Since these fields are homogeneous, symmetric and
diagonal, the reduced phase space is 6 dimensional. It is convenient
to coordinatize it with just six numbers, $C_I, P^I$, with
$I=1,2,3$:
\be C_1 := \int_{{}^3\!M}\, C_1{}^1;\quad
 P^1 :=\int_{{}^3\!M}\, P_1{}^1; \quad {\rm etc} \ee
where the integrals are well defined because we have completely
fixed the internal gauge, in that gauge the integrands are all
densities of weight 1, and ${}^3\!M$ is compact. From now on we will
focus on the description of $\mathcal{P}_{\rm red}$ in terms of
$C_I$ and $P^I$.

The symplectic structure on $\mathcal{P}_{\rm red}$ is given by the
Poisson brackets:%
\footnote{Note that, thanks to the integrals in the definitions
of $C_I$ and $P^I$, the delta-distributions on the right hand
side of truncated Poisson brackets (\ref{TPB1}) - (\ref{TPB2})
on $\mathcal{P}$ have now disappeared. To write the truncated
constraints (\ref{S2}), (\ref{V2}) in terms of $C_I, P^I$, one
first fixes a nowhere vanishing scalar density $S_1$ of weight
1 (i.e., a 3-form; see Appendix A). One then multiplies these
constraints $(S_{1})^{-2}$ to obtain constraints with density
weight zero. Finally, by noting that $C_1 = (C_1{}^1 S_{-1})\,
V_o$, etc,  where $V_o$ is the volume of ${}^3\!M$ with respect
$(S_{1})$, one obtains the equations of motion for $C_I, P^I$
given below.}
\be
    \{P^I,P^J\} = \{C_I,C_J\} = 0 \quad {\rm and} \quad
    \{P^I, C_J\} = 2\delta^I_J\, C_J\, .
\ee
The scalar or Hamiltonian constraint
\be \label{BKLHam} \frac{1}{2}  C^2 - C_IC^I + \frac{1}{2} P^2 -
P_IP^I=0 \ee
now generates the equations of motion via Poisson brackets:
 \ba
    \label{kadot} \dot{P}_I = {N} C_I \big(C- 2 C_I \big) \\
    \label{gadot} \dot{C}_I = -{N} C_I \big( P - 2 P_I \big)
\ea
Here and in what follows we use the summation convention also for
the indices $I,J$ and have set
\be  P = P_1+P_2+P_3   \quad  {\rm and} \quad  C = C_1+C_2+C_3 \, .
\ee
As a side remark, we note that $C_I=0$ is a fixed point of our
system for each $C_I$, whence the sign of each $C_I$ along any
dynamical trajectory is fixed by the initial conditions. Therefore,
away from the `planes' $C_I =0$, we can, if we wish, perform a
change of variables to $X_I= \ln|C_I|/2$ and work with the
canonically conjugate pair $(X^I, P_I)$. However, in what follows,
we will continue to work with $(C_I, P^I)$.

Finally, recall that in the BKL conjecture `the only matter that
matters' is a scalar field. Let us therefore extend our
gravitational reduced phase space to include a massless scalar field
$\phi$. Denote the conjugate momentum by $\pi$ so that $\{\phi,\,
\pi\} =1$. Then on this extended reduced phase space
$\bar{\mathcal{P}}_{\rm red}$ the Hamiltonian constraint is given by
\be \label{BKLHam2}\frac{1}{2}  C^2 - C_I C^I + \frac{1}{2} P^2 -
P_IP^I - \f{\pi^2}{2}  =0 \ee
The equations for $\dot{P}$ and $\dot{C}$ are still given by
(\ref{kadot}) and (\ref{gadot}) while those of the scalar field are
simply $\dot\phi = \pi$ and $\dot\pi = 0$.

\subsection{Dynamics}
\label{s5.3}

The Hamiltonian flow in $\bar{\mathcal{P}}_{\rm red}$ fully captures
the gauge invariant properties of the truncated dynamics of fields
at any one fixed spatial point on ${}^3\!M$. Let us therefore focus
on this Hamiltonian system. Although the basic constraint and
evolution equations on $\bar{\mathcal{P}}_{\rm red}$ are just ODEs,
they have a rich structure; indeed they incorporate the dynamics of
all Bianchi Type A models. Since the analysis of Bianchi IX is
already quite complicated and required considerable effort
\cite{Ringstrom,Ringstrom2}, we will follow the strategy used in
\cite{UEWE} and analyze implications of the reduced equations near
fixed points.

There are two sets of fixed points of the dynamics, i.e., points at
which $\dot{C_I}=\dot{P_I}=0$:
\bnum
\item $C_1=C_2$, $C_3=0$, $P_1=P_2$, $P_3=0$, and $\pi=0$
\item $C_I = 0$ and $ P_IP^I -\frac{1}{2} P^2 +
    \frac{1}{2}\pi^2=0$\, . \enum
The first set of fixed points corresponds essentially to a
dimensional reduction of our theory \cite{u1} and is therefore
highly unstable. To show that our truncation captures the standard
features associated with the BKL behavior near singularities, it
will suffice to focus on the second set which, we will now show, in
fact corresponds to the Kasner solutions. One can show that the
solutions to the scalar constraint $2P_IP^I -P^2 +\pi^2 =0$ are
such that \emph{all three $P_I$ are positive or all three are
negative}. Choice of positive signs turns out to be necessary and
sufficient for the singularity to appear at $t= +\infty$ as per our
previous conventions.

Let us return for a moment to the homogeneous phase space
$\mathcal{P}_{\rm hom}$ and set lapse $N = S_{-1}$, the
fiducial scalar density for which $S_{-1} P_{ij}$ is
homogeneous, diagonal, with entries $P_I$. We can then solve
the evolution equation (\ref{Edot}) for the triad $E^a_i(t)$ in
terms of $P_I$. Finally let us set
\be p_I = 1-\f{2P_I}{P}  \qquad {\rm and} \qquad \tau = e^{-Pt/2} \, .
\ee
Then the space-time metric computed from $E^a_i(t)$ is given by
\be \label{KasnerMetric} ds^2 = -d\tau^2 + \tau^{2p_1}dx_1^2 +
\tau^{2p_2}dx_2^2 +\tau^{2p_3}dx_3^2 \,  \ee
so that the singularity lies at $\tau=0$ (or $t=\infty$). By
definition, the constants $p_i$ satisfy
\be p_1 +p_2+ p_3 =1\ee
and the Hamiltonian constraint
\be 2P_I P^I- P^2+ \pi^2=0 \ee
on $P_I$ translates to the familiar quadratic Kasner constraint
\be p_1^2 +p_2^2 +p_3^2 =1-p_{\phi}^2\quad {\rm where}\quad p_\phi^2
= \f{2\pi^2}{P^2}\,. \ee
For each value of $p_\phi<1$, these constraints on the $p_i$ define
a 1-parameter family of solutions, the intersection of a plane with
a 2-sphere. One can check that if $p_\phi^2 >1/2$ all the $p_i$ are
positive while if $p_\phi^2<1/2$ solutions exist only if one of the
$p_i$ is negative. We will now show that this distinction plays the
key role for the stability of the solution.

Let us now move away slightly from a Kasner fixed point $(P_I,
C_I)$ and consider the Hamiltonian trajectory through the new
point $(P'_I, C'_I)$:
\be P'_I=P_I + \delta P_I, \quad\quad C'_I = C_I + \delta C_I \, .
\ee
Then, the evolution equations for the perturbations are of the form
\be \dot{(\delta P_I)} = \mathcal{O}(\delta P^2) \quad {\rm and}
\quad \dot{(\delta C_I)} = -N \delta C_I (P - 2P_I) +
\mathcal{O}(\delta C \delta P) \ee
For definiteness, let us set $I=1$. Then $P- 2 P_1 = p_1 P$ and
similarly for $I=2,3$. Now $P$ is positive since all three $P_I$ are
positive. Therefore, if all $p_i$ are positive (i.e. if  $p_\phi^2 >
1/2$), the evolution equation for $\delta C_I$ is of the type
$\dot{(\delta C_I)} = {\hbox{\rm (negative definite
quantity)}}\,\times\, \delta C_I$, whence the perturbation will
decay, implying stability. In terms of the canonical variables
describing the scalar field, this occurs when the scalar field is
large: $4\pi^2 > P^2$. This stability is in accordance with the
Andersson-Rendall results \cite{ar} on approach to space-like
singularity in presence of a massless scalar field in full general
relativity.

\begin{figure}[ht]
\includegraphics [width=380pt] {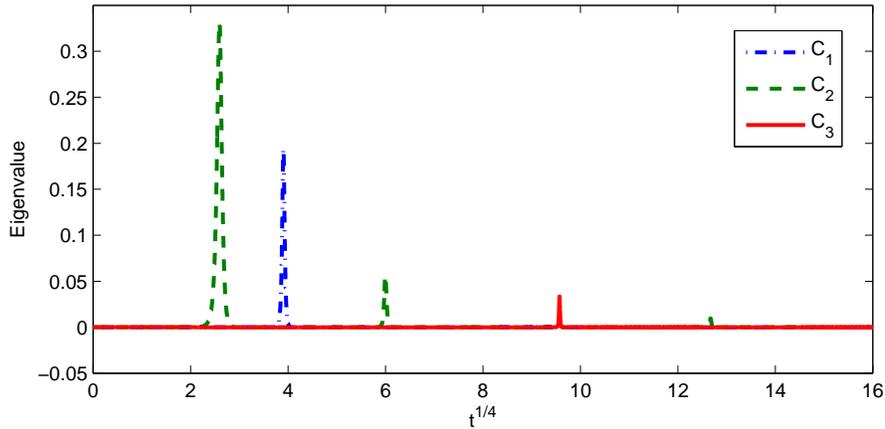}
\caption{Evolution of each of the three $C_I$ in vacuum, starting
from a point near the Kasner fixed point surface. Initial data
is $C_1=1 \times 10^{-7}$ ,\, $C_2=2 \times 10^{-7}$,\, $C_3=2.2
\times 10^{-7}$, $P_1=0.4$,\, $P_2=0.8$,\, $P_3=0.0686$ ($C_1$
in blue, $C_2$ in green, $C_3$ in red). Since none of the initial
$C_I$ vanish, as expected from analytical considerations, there is
a series of separate Taub transitions between Kasner states. Time
has been rescaled by a power of 1/4 to allow multiple transitions
to be shown on a single plot.} \label{clong}
\end{figure}
\begin{figure}[ht]
\includegraphics [width=380pt] {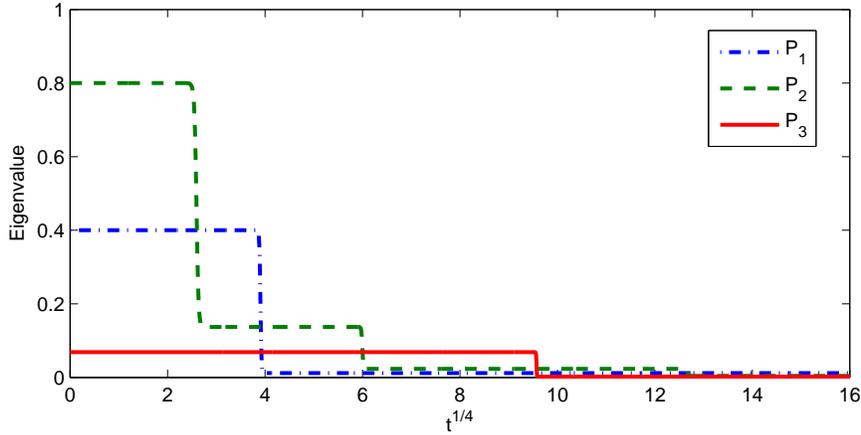}
\caption{Evolution of each of the three $P_I$ in vacuum, starting
from a point near the Kasner fixed point surface, with the same
initial data as in Fig. \ref{clong} ($P_1$ in blue, $P_2$ in green,
$P_3$ in red).
The largest eigenvalue, $P_2$, transits first. After this transition,
$P_1$ becomes the largest eigenvalue, now making $C_1$ unstable.
In time all three $P_I$ tend to zero. In terms of parameters $p_i$
used in the Kasner metric (\ref{KasnerMetric}), the initially expanding
direction $p_2$ starts contracting at the end of the transition
and initially contracting $p_1$ starts expanding.}
\label{plong}
\end{figure}

Let us now consider the complementary case where $p_\phi^2 <
1/2$. By the above reasoning, now $(P -2P_I)$ is negative for
some $I$. For definiteness, let us take $P_1$ to be the largest
of the $P_I$'s initially so that $(P - 2P_1)$ is negative which
implies that $C_1$ will grow and we have instability. In this
case, we cannot use perturbative analysis for the pair
$C_1,P_1$; it is necessary to keep all order terms in $C_1$ and
$P_1$. For simplicity, let us set $C_2=C_3=0$ initially. Then
values of $C_2, P_2, C_3, P_3$ will not change during evolution
and equations for $C_1,P_1$ simplify,
\ba \label{Stability}
\dot{P_1} &=& - N C_1^2 \\
\dot{C_1} &=& - N C_1 (P_2 + P_3 - P_1) = -N C_1 P p_1\, , \ea
which can be solved exactly to obtain
\ba \label{cptrans} P_1(t) &=& P_2 + P_3 - 2 \sqrt{P_2 P_3}\,\,
\tanh\big(2\sqrt{P_2 P_3} N (t-t_o)\big) \nonumber\\
C_1(t) &=& \pm 2 \sqrt{P_2 P_3}\,\, {\rm sech} \big(2 \sqrt{P_2
P_3} N(t-t_o)\big)\, .  \ea
These are the Bianchi II solutions written in our variables. Here
$C_1$, the unstable variable, rapidly increases and then decays to
zero.  During that time the $P_1$ transitions between one Kasner
solution to another. In the asymptotic limits we have:
\ba \label{pout}
P_1(-\infty) &=& P_2 + P_3 + 2 \sqrt{P_2 P_3} =
(\sqrt{P_2}+ \sqrt{P_3})^2 \nonumber \\
P_1(+\infty) &=& P_2 + P_3 - 2 \sqrt{P_2 P_3} =
 (\sqrt{P_2} -\sqrt{P_3})^2 \, .\ea
(In practice the asymptotic limits are achieved quickly, thanks
to the hyperbolic functions of time.) The result of the
transition is that $P_1$, which was originally was the largest
of the three $P_I$, has transitioned to a lower value. By a
change of variables to the $p_i$ used in (\ref{KasnerMetric})
it is apparent that the eigenvalue corresponding to the
negative exponent $p_i$ is the one which has transitioned, and
is positive at the end of the transition. Since the singularity
lies at $\tau=0$, this means that the initially expanding
direction now contracts, and one of the two contracting
directions now expands. Indeed, (\ref{pout}) is precisely the
`u-map' in $p_i$ variables.

\begin{figure}[ht]
\includegraphics [width=380pt] {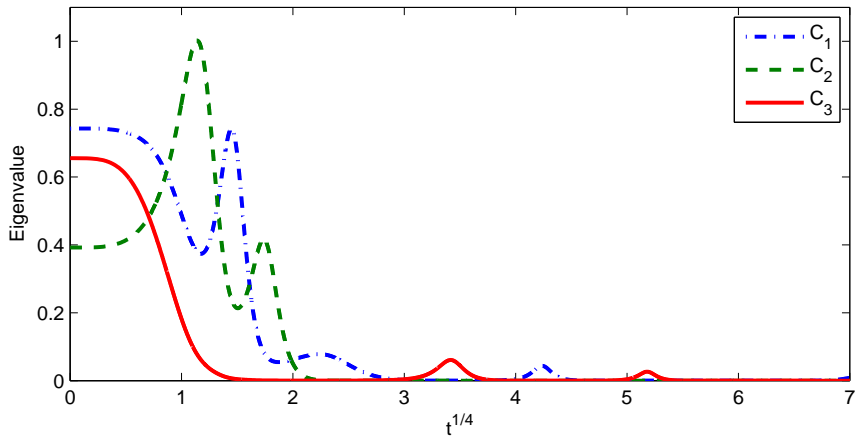}
\caption{Evolution of each of the three $C_I$ in vacuum, starting
from a point away from the Kasner fixed point surface. Initial data
are $C_1= 0.7431$,\, $C_2= 0.3922$,\, $C_3=0.6555$,
$P_1=0.1712$,\, $P_2=0.0.7060$, \, $P_3= -0.3140$\, ($C_1$ in blue, $C_2$
in green, $C_3$ in red). Even though we start out far from the Kasner
surface (where all $C_I$ vanish), dynamics drives the state to the
Kasner surface. Again, time has been rescaled by a power of 1/4 to allow
multiple transitions to be shown on a single plot.} \label{cgeneric}
\end{figure}

In this analysis we have made the simplification that initially
$C_2= C_3 =0$. If one starts from a generic point in the
vicinity of the Kasner fixed point set and still with $P_1$ as
the largest of the three $P_I$ initially, there would again be
a transition of the type (\ref{cptrans}). But as $P_1$
decreases, after a \emph{finite} time either $P_2$ or $P_3$
will now be the largest eigenvalue and making the corresponding
$C_I$ unstable. That pair will then evolve according to
(\ref{cptrans}). This general scenario was borne out in a large
class of simulations of the reduced equations of motion. Figs
\ref{clong} and \ref{plong} illustrate this dynamical behavior
for generic initial data near the Kasner surface. The Taub
transitions are easy to see in Fig \ref{plong}: even though
none of the $C_I$ are initially zero, the Taub transitions are
well described by the analytical expressions (\ref{cptrans}).
Fig \ref{cgeneric} %and \ref{pgeneric}
illustrates the dynamical behavior in cases where the initial
data is quite far from the Kasner surface. Note that even in
this case, the $C_I$ decrease in time so that, although we
start far away from the Kasner surface, dynamics drives the
state to the Kasner surface.

We can also draw some lessons for the full theory from this behavior
of the truncated system. Recall that the dynamical trajectories
discussed above can be thought of as representing the evolution of
fields at a fixed spatial point. Let us therefore return to
${}^3\!M$ and consider fields $C_I(x), P_I(x)$. Now, generically, we
will encounter a point $x_0$ where the $C^I$ all vanish, while being
non-zero is the neighborhood of the point. As we noted in section
\ref{s5.2}, the sign of $C_I$ is preserved throughout the evolution.
This is in particular true during the Taub transitions where the
magnitude of $C_I$ grows. Therefore, on `one side' of $x_0$, a $C_I$
will positive and increasing in magnitude, while on the `other side'
it will be negative and increasing in magnitude. Therefore its
derivative will increase rapidly. Similarly the under Taub
transitions the values of $P_I(x)$ will change from those of one
Kasner solution to another \emph{except at the point} $x_0$. Again,
this dynamics will generate a large derivative at $x_0$. Thus,
analysis of the reduced system suggests that \emph{spikes} will
occur in the full system (\ref{Cdot1}) - (\ref{Pdot1}). As is well
known, these spikes were found in numerical simulations and, more
recently, also in analytical treatments \cite{rw,WCL}. Whenever
spikes appear, the key assumption underlying BKL truncation is
brought to question because the spatial derivatives are large at the
spikes. The key issue for the BKL conjecture ---and for the
application to quantum gravity we proposed in section \ref{s1}--- is
whether the time derivatives still dominate generically, as they do
in examples.

Let us summarize. Using analytical and numerical methods we showed
that there exists a well defined subspace $\mathcal{P}_{\rm hom}$ of
the full phase space $\mathcal{P}$ which exhibits exactly the
properties expected in the BKL conjecture. Our procedure to arrive
at the reduced system is more direct than those available in the
literature. In \cite{UEWE}, for example, elimination of the
off-diagonal components and the anti-symmetric parts of $C_{ij}$
involves an additional assumption, beyond ignoring spatial
derivatives in favor of time derivatives: these quantities are
identified as part of the `stable subset' (variables that are
expected to decay rapidly as the singularity is approached) and then
set to zero to obtain the truncated equations. In our treatment, on
the other hand, the fact that the antisymmetric part of $C_{ij}$ is
negligible is directly implied by the assumption that the $\D_i$
derivatives are negligible and the constraints \emph{imply} that the
variables $C_{ij}$ and $P_{ij}$ can be simultaneously diagonalized
which, furthermore, completely fixes the gauge. Thus, our
Hamiltonian framework naturally led to a diagonal gauge, enabling us
to quickly zero-in on the essential variables and eliminating the
need to keep track of the dynamics of extraneous variables involving
frame rotations \cite{UEWE,ddb}. Finally, the framework easily led
us to the Mixmaster behavior ---a series of Bianchi I phases
interspersed by Bianchi II transitions. We recovered the `u-map' for
these transitions, and observed the behavior expected from the
Andersson and Rendall analysis \cite{ar} when a scalar field of
large enough magnitude is introduced.

\section{Discussion}
\label{s6}

We began with the Hamiltonian formulation of general relativity
underlying LQG where the basic fields are spatial triads $E^a_i$
with density weight 1, spin connections $\Gamma_a^i$ they determine,
and extrinsic curvatures $K_a^i$. Based on examples that have been
studied analytically and numerically, it seems reasonable to expect
that the determinant $q$ of the spatial metric $q_{ab}$ would vanish
and the trace $K$ of the extrinsic curvature would diverge at
space-like singularities. (This expectation is in particular borne
out in the numerical simulations of G2 space-times \cite{WCL1}.) One
can therefore hope to obtain quantities which remain well-defined at
the singularity either by multiplying the natural geometric fields
by suitable powers of $q$ or dividing them by suitable powers of
$K$. In the commonly used framework due to Uggla et al
\cite{UEWE,Uggla,dg1}, one chooses to divide by $K$. One first
introduces the so-called Hubble normalized triad $K^{-1} e^a{}_i$ by
rescaling the orthonormal triad $e^a_i$ by $K^{-1}$, and then
constructs a set of Hubble normalized fields by contracting
$\Gamma_a{}^j$ and $K_a{}^j$ with $K^{-1} e^a{}_i$. These fields are
expected to have regular limit at the space-like singularity.
Einstein's equations expressed in terms of them naturally suggests a
truncation and the truncated system successfully describes the
expected oscillatory BKL behavior. The resulting form of the BKL
conjecture is supported by numerical evolutions of full general
relativity carried out by Garfinkle \cite{Garfinkle}. However,
because there is no underlying Hamiltonian framework, this approach
does not easily lend itself to non-perturbative quantization. Even
if such a framework were to be constructed, because of the presence
of the $K^{-1}$ factor, it would be difficult to introduce quantum
operators corresponding to the Hubble rescaled fields.

Motivated by quantum considerations, we adopted the complementary
strategy of multiplying geometrical fields by $\sqrt{q}$. The LQG
Hamiltonian formulation we began with already features a density
weighted triad with \emph{exactly} the desired property: $E^a{}_i =
\sqrt{q} e^a_i$. Since $\sqrt{q}$ is expected to vanish at the
singularity, one can hope to use $E^a_i$ in place of the Hubble
normalized $K^{-1} e^a_i$ to construct a new set of fields to
formulate the BKL conjecture. Indeed, (modulo trace terms) our basic
variables $C_i{}^j$ and $P_i{}^j$ were obtained simply by contacting
the spatial indices of $\Gamma_a{}^j$ and $K_a^j$ by $E^a_i$.
Furthermore, because $E^a_i$ vanishes in the limit, the operator
$D_i:= E^a_i D_a$ provided a key tool in the formulation of the BKL
conjecture: asymptotically, $D_i C_j{}^k$ and $D_i P_j{}^k$ should
become `negligible' relative to $C_j{}^k$ and $P_j{}^k$. Now, in
exact general relativity, time derivatives of $C_i{}^j$ and
$P_i{}^j$ can be expressed in terms of their $D_i$ derivatives,
purely algebraic (and at most quadratic) combinations of $C_i{}^j$
and $P_i{}^j$, the lapse $N$ and its $D_i$ derivatives (see
(\ref{S1})--(\ref{Ddot1})). Therefore, if in the limit the $D_i$
derivatives of the basic fields become negligible compared to the
fields themselves, we are naturally led to conclude that time
derivatives would dominate the spatial derivatives. This chain of
argument led to our formulation of the BKL conjecture.

This rather simple idea depends on the fact that the structure of
Einstein's equations has an interesting and unanticipated feature:
as we saw in section \ref{s3}, once the triplet $C_i{}^j, P_i{}^j,
D_i$ is constructed from the triad $E^a_i$ and the extrinsic
curvature $K_a^i$ on an \emph{initial slice}, the constraint and
evolution equations can be expressed entirely in terms of the
triplet. Given a solution to these equations, the spatial triad
$E^a_i$ (and hence the metric $q_{ab}$) can be recovered at the end
simply by solving a total differential equation (\ref{Edot}). This
is a surprising and potentially deep property of Einstein's
equation. It played essential role in our formulation of the BKL
conjecture and could well capture the primary reason behind the BKL
behavior observed in examples and numerical simulations.

Since our framework is developed systematically from a Hamiltonian
theory, its BKL truncation naturally led to a truncated phase space.
The specific truncation used has an important property: The
truncated constraint and evolution equations on the truncated phase
space coincide with the truncation of full equations on the full
phase space. On the truncated phase space we could solve and
gauge-fix the Gauss and vector constraints to obtain a simple
Hamiltonian system (which encompasses all Bianchi type A models).
Solutions to this system were explored both analytically and
numerically. We showed that they exhibit the Bianchi I behavior, the
Bianchi II transitions and spikes as in the analysis of symmetry
reduced models \cite{pretlim} and numerical investigations of full
general relativity \cite{dg1}. Therefore, as explained in section
\ref{s1}, an appropriate quantization of the truncated system, e.g.,
a la loop quantum cosmology, could go a long way toward
understanding the fate of generic space-like singularities in
quantum gravity.

In sections \ref{s3}, \ref{s5.1} and \ref{s5.2}, we restricted
ourselves to vacuum equations. The addition of a massless scalar
field is straightforward and was carried out in the reduced phase
space framework in section \ref{s5.3}. If the energy density in the
scalar field is small, one again has Bianchi II transitions and
spikes. However, once the energy density exceeds a critical value,
these disappear and the asymptotic dynamics at any spatial point is
described just by the Bianchi I model with a scalar field without
transitions. Thus, our truncated system faithfully captures the main
features generally expected from the analysis of Andersson and
Rendall \cite{ar} in full general relativity coupled to a massless
scalar field or stiff fluid. Thus, although the initial motivation
came from quantum considerations, \emph{our formulation of the BKL
conjecture, and the form of the field equations both in the full and
truncated versions, should be useful also in the analytical and
numerical investigations of singularities in classical general
relativity.}

We will conclude with a discussion comparing our approach with that
of Uggla, Ellis, Wainwright and Elst (UEWE) (\cite{UEWE}). The
Hubble normalized variable used in their formulation of field
equations are give by
\ba \label{hn1} \Sigma_{ij} &=& 3K^{-1} e^a_{(i} K_{|a|j)} - K^{-1}
e^a_k K_a^k \delta_{ij} \\
\label{hn2} N_{ij} &=& -3K^{-1} e^a_{(i} \Gamma_{|a|j)} +
3K^{-1} e^a_k \Gamma_a^k \delta_{ij} \\
\label{hn3} A_i &=& -\ep_i^{\;jk} 3K^{-1} e^a_j \Gamma_a^k \\
\label{hn4} \partial_i &=& 3K^{-1} e^a_i \partial_a \, .\ea
These variables are especially useful because they are scale
invariant: they are unchanged under a constant rescaling of the
space-time metric. Because of this property and because of the
`regulating' factor $K^{-1}$ in their expressions, it is hoped that
in the limit as one approaches the space-like singularity, these
variables will remain finite \cite{Uggla} and their $\partial_i$
derivative will become negligible.

We began with quite a different motivation and our focus was on
constructing a Hamiltonian framework rather than on differential
equations. Since our emphasis was on constructing phase space
variables that can be readily promoted to well-defined quantum
operators, from the start we avoided the use of factors such as
$1/K$. As a result, our basic variables $C_i{}^j$ and $P_i{}^j$ are
\emph{not} scale invariant. Could we have made a different choice
which is also well suited for quantization and at the same time
enjoyed scale invariance? The answer is in the negative for the
following reason. Under constant conformal rescalings $g_{ab} \to
\lambda^2 g_{ab}$ of the space-time metric, we have $E^a_i \to
\lambda^2 E^a_i$,\, $\Gamma_a^i \to \Gamma_a^i$, and $K_a^i \to
K_a^i$. Now, in the analysis of approach to singularity, scale
invariant quantities are directly useful only if they are space
\emph{scalars} and it is \emph{not} possible to construct scale
invariant scalars using just sums of products of these fields, i.e.,
without introducing fields such as $K^{-1}$ for which it is
difficult to construct quantum operators. Even if one introduces
additional non-dynamical fields, such as fiducial frames to
construct scalars, for natural choices of these frames, scale
invariant components of fields such as $K_a^i, \Gamma_a^i$ typically
diverge at the singularity. Thus, with our motivation, it does not
seem possible to demand scale invariance of the basic variables that
are to feature in the BKL conjecture.

Our viewpoint is that the most important feature of the Hubble
normalized variables is that although the orthonormal triad $e^a_i$
typically diverges as one approaches a space-like singularity, $K$
diverges even faster, making the combination $K^{-1}e^a_i$ go to
zero at the singularity. Furthermore, it goes to zero at a
sufficient rate for its contraction with $K_a^i,\, \Gamma_a^i$ and
$\partial_a$ in (\ref{hn1}) --- (\ref{hn4}) to tame the a priori
divergent behavior of these fields. Instead of dividing the
orthonormal triad $e^a_i$ by $K$ which one expects to diverge at the
singularity, our strategy was to multiply it by the volume element
$\sqrt{q}$ which, in examples, goes to zero at the singularity. This
difference persists also in the treatment of the lapse. The UEWE
framework assumes that the (scalar) lapse $\bar{N}$ is such that
$\bar{N}K$ admits a limit $\underbar{N}$ while we assume that the
density weighted lapse $N = (\sqrt{q})^{-1}\, \bar{N}$ admits a
well-defined limit at the singularity. Thus, in both cases, the
standard scalar lapse $\bar{N}$ goes to zero so the singularity lies
at $t=\infty$.

The key scale invariant UEWE variables $(N_{ij}, \Sigma_{ij})$
---which are expected to be well behaved at the singularity--- are
related to our $(C_{ij}, K_{ij})$ via
\ba N_{ij} = 6 P^{-1}\, C_{(ij)}\quad &{\rm and}& \quad \Sigma_{ij}
= -6 P^{-1}\, P_{(ij)} + 2 \delta_{ij}\, , \quad {\rm or}, \\
C_{(ij)} = - \f{K\sqrt{q}}{3} N_{ij} \quad &{\rm and}& \quad
P_{(ij)} = \f{K\sqrt{q}}{3}\, (\Sigma_{ij} - 2 \delta_{ij})\,
\ea
and the two sets of lapse fields are related by:
\be \underbar{N}= K \sqrt{q}\, N \, .\ee
If one focuses only on the structure of differential equations near
space-like singularities, the two reduced systems would in essence
be equivalent if $K\sqrt{q}$ admits a finite, nowhere vanishing
limit at the singularity. This condition holds for Bianchi I models
and also Bianchi II which describe the transitions between Bianchi I
epochs. In fact in the Bianchi I model, $\sqrt{q}\, K =1$ and our
density weighted triad has the \emph{same} dependence on proper time
as the Hubble normalized triad. Thus, although the motivations,
starting points and procedures used in the two frameworks are quite
different, surprisingly, in the end the basic variables and
equations are closely related.

\section*{Acknowledgments}

We would like to thank Woei-Chet Lim, Alan Rendall, Yuxi Zheng and
especially David Garfinkle and Claes Uggla  for discussions. This
work was supported in part by the NSF grant PHY0854743 and the
Eberly and Frymoyer research funds of Penn State.

\begin{appendix}

\section{Densities}
\label{a1}

Since the basic variables that feature in our formulation of the BKL
conjecture are scalars on ${}^3\!M$ of density weight 1, in this
Appendix we briefly recall a coordinate independent framework for
describe densities. The underlying idea is due to Wheeler and the
detailed framework was developed by Geroch (see, e.g., \cite{rg}).
This framework goes hand in had with Penrose's abstract index
notation \cite{rp,ahm}. Because the primary application in this
paper is to our fields $C_i{}^j, P_i{}^j$ on ${}^3\!M$ we will focus
on scalar densities on 3-manifolds. But generalization to tensor
densities on n-manifolds is straightforward.

Fix an oriented 3-manifold ${}^3\!M$ and fix a orientation thereon.
Denote by $\CE$ the space of smooth, positively oriented, no-where
vanishing, totally skew tensor fields $e^{abc}$ on ${}^3\!M$.
Clearly, given any two elements $e^{abc}$ and $e^\prime{}^{abc}$ in
$\CE$, there exists a (strictly) positive function $\alpha$ such
that $e^\prime{}^{abc}= \alpha e^{abc}$. This fact will be used
repeatedly.

In this paper, a scalar density $S_n$ of weight $n$ is a map from
$\CE$ to the space of (real valued) smooth functions on ${}^3\!M$:
$e \rightarrow S_n(e)$, such that:
\be S_n(e^\prime) = \alpha^n\, S_n(e)  \ee
Here $n$ can be any real number but in most applications in general
relativity it is an integer. (In quantum mechanics, on the other
hand, states are (complex-valued) densities of weight $1/2$ on the
configuration space \cite{rg}.) Since $\C_i{}^j, P_i{}^j$ have
density weight $1$, let us make a short detour to discuss the case
$n=1$. Fix any 3-form $s_{abc}$ on ${}^3\!M$. It determines a
canonical scalar density of weight $1$:
\be S_1(e) : = s_{abc}\, e^{abc} \, . \ee
Conversely, since $S_1$ is a linear mapping from $\CE$ to
smooth functions, it determines a canonical 3-form $s_{abc}$.
Thus, our basic variables could also be taken to be 3-forms
$C^{ij}{}_{abc}, \, P^{ij}{}_{abc}$ on ${}^3\!M$ which take
values in second rank tensors in the internal space. The
standard ADM phase space of general relativity can be similarly
coordinatized by positive definite metrics $q_{ab}$ and tensor
fields $P^{ab}_{cde}$ which are symmetric in $a,b$ and totally
skew in $c,d,e$ \cite{ag,am}. Finally note that every metric
$q_{ab}$ determines a canonical volume 3-form $\epsilon_{abc}$
which has positive orientation and satisfies
$\epsilon_{abc}\epsilon_{def}\, q^{ad} q^{be} q^{cf} = {\rm
sgn}(q)\,\, 3!$. Therefore it also determines a canonical
scalar density $\sqrt{q}$ of weight $1$, called the
\emph{square root of the determinant} of $q_{ab}$:\,
$\sqrt{q}(e) := \epsilon_{abc}e^{abc}$ for all $e\in \CE$.

This definition can be extended to density weighted tensor
fields in an obvious fashion. Note that every ${}^3\!M$ carries
a natural totally skew tensor density $\eta^{abc}$ of weight
$1$, called the Levi-Civita density:
\be \eta^{abc} (e) = e^{abc} \qquad \forall\, e \in \CE \ee
Given any metric $q_{ab}$ on ${}^3\!M$, the square root of its
determinant, $\sqrt{q}$, can also be expressed as $\sqrt{q} =
\eta^{abc}\, \epsilon_{abc}$.

Finally, given a derivative operator $D_a$ on tensor fields on
${}^3\!M$, we can extend its action on densities $S_n$ of weight $1$
in a natural manner. $D_a S_n$ is a 1-form with the same density
weight $n$, given by
\be (D_a S_n)(e) = D_a (S_n(e)) - n \lambda_a\,  S_n(e) \qquad
\forall e \in \CE \, ,\ee
where the first term on the right hand side is just the gradient of
the function $S_n(e)$ and the 1-form $\lambda_a$ is given by $D_a
e^{bcd} = \lambda_a e^{bcd}$. Therefore the action of the derivative
operator $D_i$ introduced in the main text is given by:
\be (D_i S_n)(e) = D_i(S_n(e)) - n\, (E^a_i\lambda_a)\, S_n(e)\qquad
\forall e \in \CE \, . \ee
Since the derivative operator $D_a$  we considered ignores internal
indices, this equation gives the action of $D_i$ on $C_i{}^j$ and
$P_i{}^j$ by regarding these basic fields simply as scalar densities
with weight $1$.

\section{Full Equations of Motion}
\label{a2}

In the main text we restricted the equations of motion to the case
where the shift is zero as is the Lagrange multiplier for the Gauss
constraint.  In this appendix we give the equations of motion in
full generality for both full general relativity and in our reduced
system.  The full equations of motion for $\C$ and $\Kbar$ are as
follows.
\ba &\dot{\C}^{ij} = - &\ep^{jkl}  \D_k ({N} (\f{1}{2}
\delta_l^i \Kbar-\Kbar_l^{\;\,i})) + {N} [2 \C^{(i}_{\;\;k}
\Kbar^{|k|j)} +
2\C^{[kj]} \Kbar_k^{\;\,i} - \Kbar \C^{ij}]\\
\nonumber&&+N^k \D_k \C_{ij} + \C_{ij} \D_k N^k +
\C_{ij} \ep_{klm} N^k \C^{lm}\\
\nonumber&& + ( \C_i^{\,\,k}\ep_{klj} + \C^k_{\,\,\,j}\ep_{kli} )
(\Lambda^l - N^m \C_{ml} + \frac{1}{2}\C N^l) \\
\nonumber&& + \D_i (\Lambda_j - N^k \C_{kj} + \frac{1}{2} N_j \C) -
\D_k (\Lambda^k - N^l \C_{l}^{\,\,k}+ \frac{1}{2} \C N^k) \delta_{ij}
\ea
\ba &\dot{\Kbar}^{ij} = &\ep^{jkl}  \D_k ({N} (1/2 \delta_l^i
\C-\C_l^{\;\,i}))  - \ep^{klm} \D_m ({N} \C_{kl})
\delta^{ij}+ 2\ep^{jkm} \C^{[ik]} \D_m ({N}) \\
\nonumber & & + (\D^i \D^j - \D^k \D_k \delta^{ij}) {N} + {N}
[- 2 \C^{(ik)} \C_k^{\;\,j} + \C \C^{ij} -
2 \C^{[kl]} \C_{[kl]}\delta^{ij}] \\
\nonumber &&+ N^k \D_k \P_{ij} + \P_{ij} \D_k N^k + \P_{ij}
\ep_{klm} N^k \C^{lm} \\
\nonumber && + ( \P_i^{\,\,k}\ep_{klj} + \P^k_{\,\,\,j}\ep_{kli} )
(\Lambda^l - N^m \C_{ml} + \frac{1}{2}\C N^l)
\ea

In the reduced system the derivative terms are set to zero
leading to the following equations of motion for $\C$ and
$\Kbar$.
\be \dot{\C}_{ij} =   {N}\, [2 \C_{k(i} \P^k{}_{j)} - \P \C_{ij}]
+ 2 \ep_{kl(i} \C_{j)}{}^{k} (\Lambda^l - N^m \C_{m}{}^{l} +
\frac{1}{2} \C N^l)
\ee
\be
 \dot{\P}_{ij} = N [- 2 \C_{ik} \C^k{}_j +\C \C_{ij}] +
 2 \ep_{kl(i} \P_{j)}{}^{k} (\Lambda^l - N^m \C_{m}{}^{l} +
  \frac{1}{2} \C N^l)
\ee

\end{appendix}


\begin{thebibliography}{99}


\bibitem{bkl1} V. A. Belinskii, I. M. Khalatnikov and E. M.
    Lifshitz, Oscillatory approach to a singular point in
    the relativistic cosmology, Adv. Phys. \textbf{31}, 525-573 (1970)

\bibitem{ahs1} A. Ashtekar, A. Henderson and D. Sloan, {Hamiltonian
    general relativity and the Belinskii, Khalatnikov, Lifshitz
    conjecture}, Class. Quant. Grav. \textbf{26} 052001 (2009)

\bibitem{Bianchi} L. Bianchi, Sugli spazii a tre dimensioni che
    ammettono un gruppo continuo di movimenti, Soc. Ital. Sci. Mem.
    di Mat. \textbf{11} 267 (1898)

\bibitem{Berger} B. Berger, Numerical approaches to space-time
    singularities, Living Reviews in Relativity \textbf{1}
    (2002)

\bibitem{Garf} D. Garfinkle {Numerical simulations of generic
    singuarities}, Phys. Rev. Lett. \textbf{93} 161101 (2004)

\bibitem{Moncrief} B. Berger and V. Moncrief {Numerical
    investigation of cosmological singularities}, Phys. Rev. D\textbf{48}
    4676 (1993)

\bibitem{Isenberg} B. Berger, D. Garfinkle, J. Isenberg, V. Moncrief
    and M. Weaver,    {The singularity in generic gravitational
    collapse is spacelike, local, and oscillatory},  Mod. Phys. Lett.
    A\textbf{13} 1565 (1998)

\bibitem{Weaver} M. Weaver, J. Isenberg and B. Berger {Mixmaster
    behavior in inhomogeneous cosmological spacetimes},
    Phys. Rev. Lett. \textbf{80} 2984 (1998)

\bibitem{ar} L. Anderson and A. Rendall {Quiescent
    cosmological singularities}, Commun. Math. Phys \textbf{218} 479
    (2001)

\bibitem{dhrw}T. Damour, H. Henneaux, A. Rendall and M. Weaver,
    Kasner-like behavior for subcritical Einstein matter
    systems, Ann. Henri Poincar\'e \textbf{3}, 1049-1111 (2002)

\bibitem{Berger2} B. Berger and V. Moncrief, {Numerical evidence
    that the singularity in polarized U(1) symmetric cosmologies on $T^3
    \times R$ is velocity dominated}, Phys. Rev. D \textbf{57} 7235
    (1998)

\bibitem{Berger3} B. Berger and V. Moncrief, {Exact U(1) symmetric
    cosmologies with local Mixmaster dynamics}, Phys. Rev.
    D\textbf{62}  023509 (2000)

\bibitem{dg1} D. Garfinkle, Numerical simulations of general
    gravitational singularities, Class. Quant. Grav. \textbf{24} 295
    (2007)

\bibitem{sag} R. Saotome, R. Akhoury and D. Garfinkle, Examining
    gravitational collapse with test scalar fields, Class. Quant. Grav.
    \textbf{27} 165019 (2010)

\bibitem{LQC} M. Bojowald, Loop quantum cosmology, Living
    Rev. Rel.8:11 (2005);\\
    A. Ashtekar, {Loop Quantum cosmology: An overview}, Gen.
    Rel. Grav \textbf{41} 707-741 (2009)

\bibitem{alrev} A. Ashtekar and J. Lewandowski,
    {Background independent quantum gravity: A status report}, Class.
    Quant. Grav. \textbf{21} R53-R153 (2004)

\bibitem{crbook} C.~Rovelli,{\em Quantum Gravity}, (Cambridge
    University Press, Cambridge (2004))

\bibitem{ttbook} T.~Thiemann, {\em Introduction to Modern
    Canonical Quantum General Relativity}, (Cambridge University Press,
    Cambridge, (2007))

\bibitem{mb1} M.~Bojowald, {Absence of singularity in loop quantum
    cosmology}, Phys. Rev. Lett. \textbf{86}, 5227-5230 (2001)

\bibitem{aps1} A. Ashtekar, T. Pawlowski and P. Singh, {Quantum
    nature of the big bang}, Phys. Rev. Lett. \textbf{96} 141301
    (2006)

%\bibitem{Closed} K. Vandersloot, {Loop quantum cosmology and the k =
%    -1 RW model}, Phys. Rev. D\textbf{75} 023523 (2007)

\bibitem{apsv} A. Ashtekar, T. Pawlowski, P. Singh and K.
    Vandersloot, {Loop quantum cosmology of k=1 FRW models}, Phys. Rev.
    D\textbf{75} 024035 (2007)

\bibitem{bp} E.~Bentivegna and T.~Pawlowski, Anti-deSitter universe
    dynamics in LQC,  Phys. Rev. D77, 124025 (2008)

\bibitem{ps} P. Singh {Are loop quantum cosmos never
    singular?} Class. Quant. Grav. \textbf{26} 125005 (2009)

\bibitem{EdB1} A. Ashtekar and E. Wilson-Ewing {Loop quantum
    cosmology of Bianchi I models}, Phys. Rev. D\textbf{79} 083535
    (2009)

\bibitem{EdB2} A. Ashtekar and E. Wilson-Ewing {Loop quantum
    cosmology of Bianchi type II models} Phys. Rev. D\textbf{80}
    123532 (2009)

\bibitem{EdB9} E. Wilson-Ewing, Loop quantum cosmology of Bianchi
    type IX models, Phys. Rev. D\textbf{82} 043508 (2010)

\bibitem{Gowdy} G. Mena Marugan and M. Martin-Benito, {Hybrid
    quantum cosmology: Combining loop and Fock quantizations}, Int. J. Mod.
    Phys. A\textbf{24} 2820 (2009)

\bibitem{UEWE} C. Uggla, H. van Elst, J. Wainwright and G.
    Ellis, {The past attractor in inhomogeneous cosmology}, Phys. Rev.
    D \textbf{68} 103502 (2003)

\bibitem{ddb} T. Damour and S. de Buyl, Describing general
    cosmological singularities in Iwasawa variables, Phys. Rev.
    D \textbf{77}, 043520 (2008)

\bibitem{aa} A. Ashtekar, New variables for classical
    and quantum gravity, Phys. Rev. Lett. \textbf{57}, 2244-2247
    (1986);\\
    A new Hamiltonian formulation of general relativity, Phys. Rev.
    D\textbf{36}, 1587-1603 (1987).

\bibitem{ADM} R. Arnowitt, S. Deser and C. W. Misner, The dynamics
    of general relativity, in {\em Gravitation: An Introduction to
    Current Research}, edited by L. Witten (Wiley, New York (1962))

\bibitem{jr} J.D~Romano, {Geometrodynamics Vs. Connection
    Dynamics}, Gen. Rel. Grav. \textbf{25} 759 (1993)

\bibitem{WCL1} W. Lim (personal communication, July 2008)

\bibitem{hur} J.M. Heinzle, C. Uggla, and N. Rohr, {The Cosmological
    Billiard Attractor}, Adv. Theor. Math. Phys. \textbf{13} 293
    (2009)

\bibitem{bh} G. Barnich and V. Hussain, Geometrical representation
    of Euclidean general relativity in the canonical formalism, Class.
    Quant. Grav. \textbf{14} 1043 (1997)

\bibitem{Ringstrom} H. Ringstrom Curvature blow up in Bianchi VIII
    and IX vacuum spacetimes, Class. Quantum Grav. \textbf{17} 713
    (2000)

\bibitem{Ringstrom2} H. Ringstrom, The Bianchi IX attractor, Annales
    Henri Poincare \textbf{2} 405 (2001)

 \bibitem{u1}C. Uggla, {The Nature of Generic Cosmological
    Singularity}, arXiv:0706.0463

\bibitem{Garfinkle} D. Garfinkle, The Nature of  gravitational
    singularities, Int. J. Mod. Phys. D \textbf{13} 2261 (2004)

\bibitem{Mixtractor} M. Heinzle and C. Uggla, A new proof of the
    Bianchi type IX attractor theorem, Class. Quant. Grav
    \textbf{26} 075015 (2009)

\bibitem{Uggla} M. Heinzle and C. Uggla, {Mixmaster: Fact and
    belief}, Class. Quant. Grav. \textbf{26} 075016 (2009)

\bibitem{Rendall2} A. Rendall, {Global dynamics of the Mixmaster
    model}, Class. Quant. Grav. \textbf{14}, 2341-2356 (1997)

\bibitem{rw} A. Rendall and M. Weaver, Manufacture of Gowdy
    space-times with spikes, Class. Quant. Grav.\textbf{18}
    2959-2976 (2001)

\bibitem{WCL} W. Lim, {New Explicit Spike Solution -- Non-local
    Component of the Generalized Mixmaster Attractor},
    Class. Quant. Grav. \textbf{25} 045014 (2008)

\bibitem{pretlim} W. Lim, L. Andersson, D. Garfinkle and F.
    Pretorius, Spikes in the Mixmaster regime of G(2) cosmologies,
    Phys. Rev. D\textbf{79} 123526 (2009)

\bibitem{rg} R. Geroch, \emph{Geometrical Quantum Mechanics},
    Lecture notes available at
    http://www.phy.syr.edu/~salgado/geroch.notes/geroch-gqm.pdf

\bibitem{rp} R. Penrose, Structure of space-time, in
    \emph{Battlle Rencontres}, edited by C. M. DeWitt and J.
    Wheeler (Bejamin, New York, (1968))

\bibitem{ahm} A. Ashtekar, G. T. Horowitz and A. Magnon, A
    Generalized tensor calculus and its applications to
    physics, Gen. Rel. Grav. \textbf{14} 411-428
    (1982)

\bibitem{ag} A. Ashtekar and  R. Geroch, Quantum theory of
    Gravitation, Rep. Prog. of Phys. \textbf{37}, 1211-1256
    (1974)

\bibitem{am} A. Ashtekar and A. Magnon, On the Symplectic
    structure of general relativity, Commun.
    Math. Phys., \textbf{86}, 55-68 (1982)

\end{thebibliography}
\end{document}